\documentclass[aps,prl,twocolumn,superscriptaddress]{revtex4-1}

\usepackage{graphicx}
\usepackage{amssymb}
\usepackage{amsmath}
\usepackage{color}

\begin{document}
	
	\title{Spin Jahn-Teller antiferromagnetism in CoTi$_2$O$_5$}
	
	\author{Franziska~K.~K.~Kirschner}
	\email{franziska.kirschner@physics.ox.ac.uk}
	\affiliation{Department of Physics, University of Oxford, Clarendon Laboratory,
		Parks Road, Oxford, OX1 3PU, United Kingdom}
	\author{Roger~D.~Johnson}
	\affiliation{Department of Physics, University of Oxford, Clarendon Laboratory,
		Parks Road, Oxford, OX1 3PU, United Kingdom}
	\author{Franz~Lang}
	\affiliation{Department of Physics, University of Oxford, Clarendon
		Laboratory, Parks Road, Oxford, OX1 3PU, United Kingdom}
	\author{Dmitry~D.~Khalyavin}
	\affiliation{ISIS Facility, Rutherford Appleton Laboratory, Harwell Oxford,
		Didcot OX11 0QX, United Kingdom}
	\author{Pascal~Manuel}
	\affiliation{ISIS Facility, Rutherford Appleton Laboratory, Harwell Oxford,
		Didcot OX11 0QX, United Kingdom}
	\author{Tom~Lancaster}
	\affiliation{Centre for Materials Physics, Durham University, Durham DH1 3LE,
		United Kingdom}
	\author{Dharmalingam Prabhakaran}
	\affiliation{Department of Physics, University of Oxford, Clarendon Laboratory,
		Parks Road, Oxford, OX1 3PU, United Kingdom}
	\author{Stephen~J.~Blundell}
	\email{stephen.blundell@physics.ox.ac.uk}
	\affiliation{Department of Physics, University of Oxford, Clarendon Laboratory,
		Parks Road, Oxford, OX1 3PU, United Kingdom}
	\date{\today}

	\begin{abstract}
		
		We have used neutron powder diffraction to solve the magnetic
		structure of orthorhombic CoTi$_2$O$_5$, showing that the long-range
		ordered state below 26 K identified in our muon-spin rotation
		experiments is antiferromagnetic with propagation vector ${\bf k}=(\pm
		\frac{1}{2}, \frac{1}{2}, 0)$ and moment of 2.72(1)$\mu_{\rm B}$ per Co$^{2+}$
		ion.
		This long range magnetic order is incompatible with the
		experimentally determined crystal structure because the imposed
		symmetry completely frustrates the exchange coupling.  We conclude
		that the magnetic transition must therefore be associated with a spin
		Jahn-Teller effect which lowers the structural symmetry and thereby
		relieves the frustration.  These results show that CoTi$_2$O$_5$ is a
		highly unusual low symmetry material exhibiting a purely spin-driven
		lattice distortion critical to the establishment of an ordered
		magnetic ground state.

	\end{abstract}
	
	\maketitle
	
	The Jahn-Teller effect is the spontaneous lowering of symmetry that
	lifts an orbital degeneracy \cite{Gehring1975} and involves a coupling
	of the orbital and lattice degrees of freedom.  In some rather rare
	cases an analogous effect can occur in which spin, rather than
	orbital, degrees of freedom play a role.  This {\it spin Jahn-Teller
		effect} has been identified in pyrochlores in which the large spin
	degeneracy in the lattice of corner-sharing tetrahedra can be relieved
	by a distortion in those tetrahedra
	\cite{Yamashita2000,Tchernyshyov2002}.
	In some cubic spinels an analogous effect can
	take place in which a tetragonal distortion relieves the frustration
	\cite{Onoda2003, Watanabe2012}. A related effect has also been
	observed near level-crossing in molecular wheels \cite{Waldmann2006, 
		Waldmann2007}.  In this Letter we demonstrate the existence of spin
	Jahn-Teller driven antiferromagnetism in CoTi$_2$O$_5$, a compound
	which has much lower symmetry than either pyrochlores or spinels,
	showing that spin-phonon coupling can induce order in a larger class
	of materials than has previously been appreciated.  The site symmetry
	of the magnetic Co$^{2+}$ (3d$^7$) ion is $m2m$ ($C_{2v}$) and so the
	orbital levels are already non-degenerate (so no longer susceptible to
	a conventional Jahn-Teller transition).  Nevertheless, we show that
	long range spin order is only permitted in the presence of the
	structural distortion that we predict to set in at $T_{\rm N}=26$~K.

	Cobalt titanates are of interest due to their numerous applications.
	Co$_2$TiO$_4$ has a complex spinel magnetic structure \cite{Ogawa1965,
		Srivastava1987, Gavoille1991, Nayak2015}, which has found uses in catalysis
	\cite{Kim2001, Zhu2009}, microwave devices \cite{Harris2009}, and Li-ion cells
	\cite{Sandhya2014}. CoTiO$_3$ has been used as a photocatalyst \cite{Ye2016},
	gas sensor \cite{Chu1999}, and also in semiconductor transistors and memory
	storage \cite{Chao2004}. CoTi$_2$O$_5$, however, is less well-studied. It is the
	only cobalt titanate to melt incongruently \cite{Brezny1969}, and its
	pseudo-brookite structure \cite{Muller-Buschbaum1983} is an entropy-stabilized
	high temperature phase \cite{Navrotsky1975} which is susceptible to
	decomposition below 1414\,K \cite{Yankin1999, Jacob2010}. Only recently has it
	become possible to synthesize high-quality single crystals of CoTi$_2$O$_5$
	\cite{Balbashov2017}. 
	
	A polycrystalline CoTi$_2$O$_5$ powder sample was prepared using high purity
	($>99.99\%$) Co$_3$O$_4$ and TiO$_2$ via the solid state reaction technique.
	Mixed powders were sintered at 1200\,$^\circ$C for 48h in air with intermediate
	grinding. After confirming the phase purity of the powder using x-ray
	diffraction, a cylindrical rod of diameter 10\,mm and length 100\,mm was
	sintered at 1250\,$^\circ$C in air for 12h.  Finally, the single crystal was
	grown in a four-mirror optical floating-zone furnace (Crystal System Inc.) in
	argon/oxygen mixed gas (90:10 ratio) atmosphere with a growth rate of 2--3mm/h.

	\begin{figure}
		\includegraphics[width=\columnwidth, clip, trim= 0.0mm 0.0mm
		0.0mm 0.0mm]{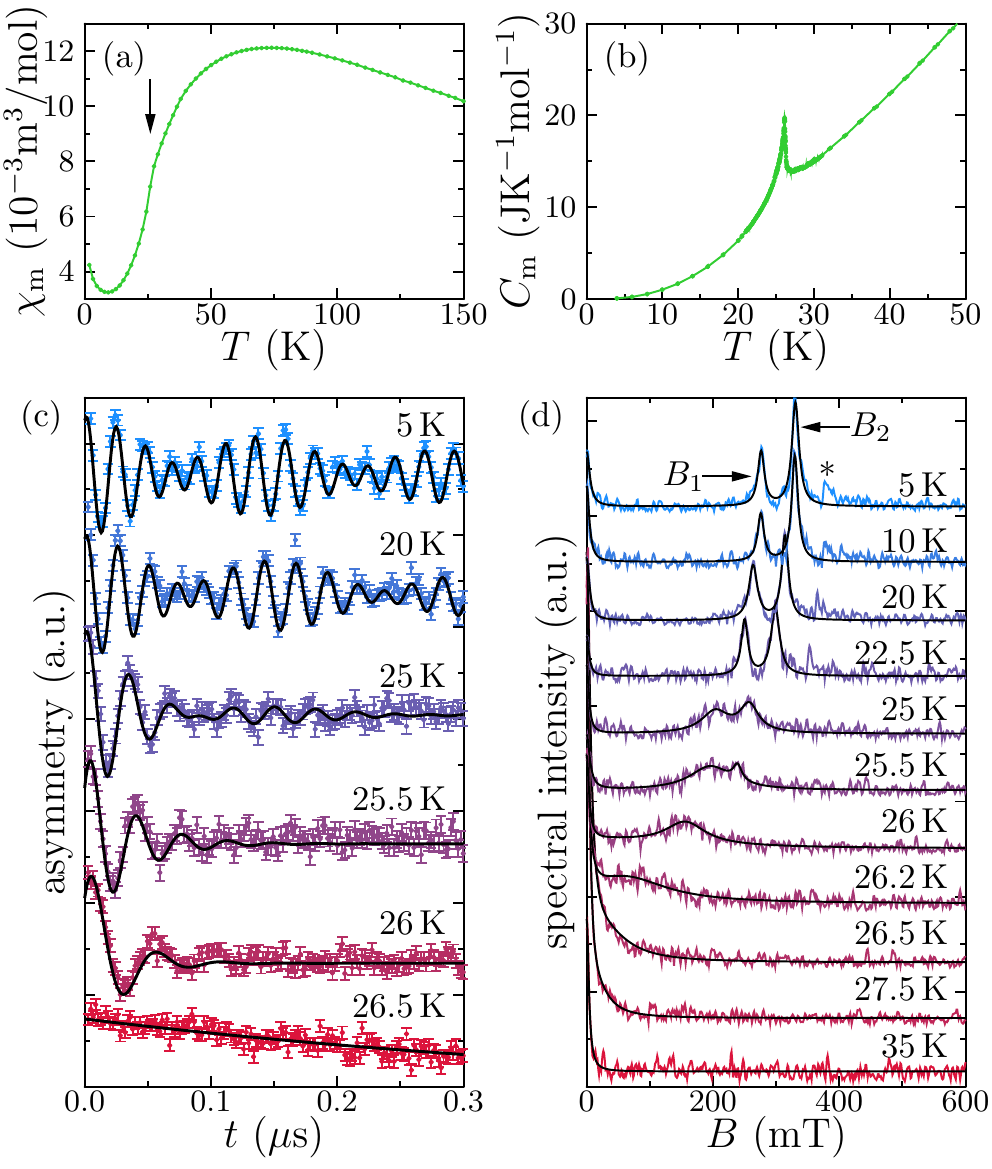}
		\caption{ \label{ZF_spec}
			(a) Magnetic susceptibility of CoTi$_2$O$_5$ measured in an applied
			field of $\mu_0 H = 0.1$\,mT. The asterisk marks a kink at $T_{\rm N}$. (b)
			Molar heat capacity. (c) ZF-$\mu$SR spectra above and below
			$T_{\rm N}$. Fits to Eq.~1 of \cite{Note2} are also
			plotted. (d) The Fourier transform of these spectra with
			fits with Eq.~\ref{eq_I}. The asterisk indicates an additional
			feature, discussed in the main text.
		}
	\end{figure}
	
	Magnetic susceptibility and heat capacity
	data are shown in  Fig.~\ref{ZF_spec}(a) and (b) respectively and are
	consistent with a magnetic transition at 26~K.  The calculated entropy
	associated with the transition is 48\% of the expected $R\ln(4)$ associated with
	the spin-only moment,
	indicative of significant correlations above $T_{\rm N}$.

	Zero field $\mu$SR (ZF-$\mu$SR) experiments \cite{Blundell1999, Yaouanc2011}
	were performed using a Quantum Continuous Flow Cryostat mounted on the general
	purpose spectrometer (GPS) at the Swiss Muon Source. All of the $\mu$SR data
	were analyzed using WiMDA \cite{Pratt2000}.
	
	ZF-$\mu$SR 
	asymmetry spectra $A(t)$ are shown in Fig.~\ref{ZF_spec}(c). At low $T$, we
	observe an oscillatory beating pattern of $A(t)$, along with two peaks in the
	Fourier transform spectra [Fig.~\ref{ZF_spec}(d)]. This is indicative of
	long-range magnetic order and two inequivalent muon stopping sites. The data can
	be fitted either in the time domain \cite{Note2} or in the field domain.
	
\begin{figure}
		\includegraphics[width=\columnwidth, clip, trim= 0.0mm 0.0mm 0.0mm
		0.0mm]{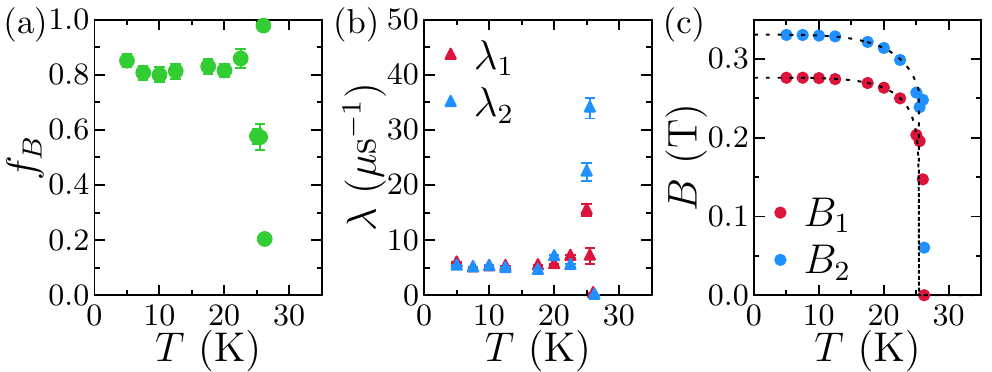}
		\caption{ \label{ZFfig}
			(a) Temperature dependence of the fraction of muons experiencing a coherent
			magnetic field. Temperature dependences of the peak width and centers, as fitted
			from Eq.~\ref{eq_I} applied to the Fourier transform of the ZF-$\mu$SR spectra,
			are shown in (b) and (c) respectively. The dotted lines in (c) show
			phenomenological fits \cite{Note2}.
		}
	\end{figure}

	Below $T_{\rm N}$, the spectral intensity $I(B)$ in the field domain can be
	modelled with a sum of three Lorentzian distributions:
	\begin{equation} \label{eq_I}
	I(B) = I_1 L(B;B_1,\lambda_1) + I_2 L(B;B_2,\lambda_2) + I_{\rm b}
	L(B;0,\lambda_{\rm b}),
	\end{equation}
	where $L(B;B_i,\lambda_i)$ is a Lorentzian distribution centred on
	$B_i$ with a width $\lambda_i/\gamma_{\mu}$ ($\gamma_{\mu} = 2 \pi
	\times 135.5$\,MHzT$^{-1}$ is the gyromagnetic ratio of the muon). The
	first two terms correspond to muons precessing in the internal fields
	of the sample; the two frequencies correspond to the two different
	dipolar fields at symmetry-inequivalent muon stopping sites. The
	Lorentzian distribution associated with each precession frequency
	indicates a small spread in the magnetic field distribution at the
	muon site, possibly due to the site disorder that has been observed
	between the Co and Ti sites in CoTi$_2$O$_5$
	\cite{Muller-Buschbaum1983}, small fluctuations of the Co moments, or
	due to muons near the boundaries of magnetic domains. The third term
	is a background term, corresponding to muons which land in the
	cryostat and sample holder, and therefore do not experience any of the
	sample's internal fields [the small width of this component,
	$\lambda_{\rm b} \approx 4$\,mT for all $T$, may be due to these
	muons experiencing a small field close to the surface of the
	sample]. The fraction of muons contributing to each peak in
	Fig.~\ref{ZF_spec}(d) $f_i$ is given by the integral under that
	peak. The total fraction of muons experiencing a non-zero field,
	$f_{B} = \left(f_1 + f_2 \right) / \left( f_1 + f_2 + f_3 \right)$, is
	plotted in Fig.~\ref{ZFfig}(a). The drop in $f_{B}$ above T$_{\rm N}$ marks the
	transition into the paramagnetic state.
	
	The fitted values from Eq.~\ref{eq_I} for the $T$ evolution of $\lambda_i$ and
	$B_i$ are presented in Figs.~\ref{ZFfig}(b) and (c) respectively. As $T$
	increases towards $T_{\rm N}$, the two peaks broaden and merge, while their
	centers move towards 0\,T as the long-range-ordered magnet transitions to the
	paramagnetic regime. The data in Fig.~\ref{ZFfig}(c) were fitted with the
	phenomenological formula $B = B_0\left(1-\left(T/T_{\rm
		N}\right)^{\alpha}\right)^{\beta}$ \cite{Note2}, giving $T_{\rm N} =
	26.0(11)$\,K for both components. We also find values of the internal fields at
	the muon sites as $T \to 0$: $B_1 = 330(3)$\,mT and $B_2 = 276(6)$\,mT. There
	appears to be an additional small feature in the data at low $T$ at $\approx
	400$\,mT (marked by the asterisk in Fig.~\ref{ZF_spec}(d)), which may arise due
	to the site disorder and is discussed below.

	Neutron powder diffraction (NPD) measurements were performed on the WISH
	time-of-flight diffractometer \cite{Chapon2011} at ISIS, the UK Neutron and Muon
	Source. A highly pure, single crystal sample was ground to a fine powder and
	loaded into a cylindrical vanadium can, which was mounted within a $^4$He
	cryostat. Data were collected with high counting statistics at 1.5 K, deep into
	the long-range ordered magnetic phase, and at 100\,K in the paramagnetic phase.
	All diffraction data were refined using FULLPROF \cite{Rodriguez-Carvajal1993}.
	
\begin{figure}
		\includegraphics[width=\columnwidth, clip, trim= 0.0mm 0.0mm 0.0mm
		0.0mm]{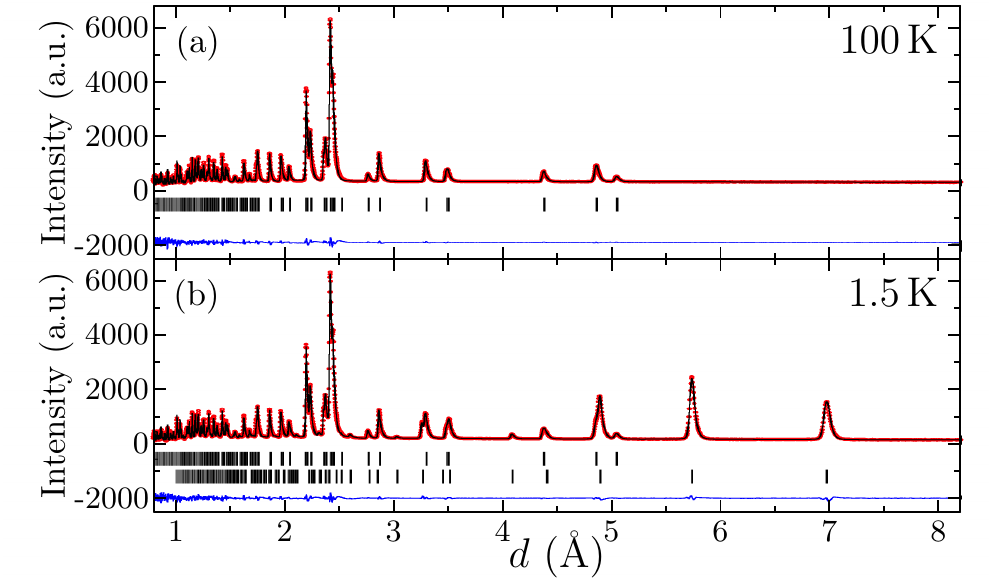}
		\caption{ \label{npd_fig}
			Neutron powder diffraction data (red points) measured in bank 2 (average
			$2\theta =58.3^{\circ}$) of the WISH diffractometer from CoTi$_2$O$_5$ in (a)
			the paramagnetic phase and (b) the antiferromagnetic phase. The fitted $Cmcm$
			nuclear model (a) and nuclear + magnetic model (b), as described in the text,
			are shown as solid black lines. The respective peak positions are shown as black
			tick marks (nuclear top, magnetic bottom). Difference patterns ($I_{\rm obs} -
			I_{\rm calc}$) are given as solid blue lines at the bottom of the panes.
		}
	\end{figure}

	NPD data collected at 100\,K (well above any anomalies in $\chi$) were fitted
	with a nuclear model based upon the published crystal structure
	\cite{Muller-Buschbaum1983}. The goodness-of-fit was excellent, the data and fit
	are shown in Fig.~\ref{npd_fig}(a), and the refined structural parameters are
	given in \cite{Note2}. There was no evidence of impurity phases in these data.
	There is a small amount of site mixing whereby 2.8\% of Co sites are occupied by
	Ti, and 1.4\% of Ti sites are occupied by Co.
	
	When compared to the NPD pattern at 100\,K, data collected at 1.5\,K showed more
	than 10 new diffraction peaks (Fig.~\ref{npd_fig}(b)). Based on bulk properties
	measurements \cite{Note2} and the results of our ZF-$\mu$SR experiments, we
	could robustly assign the origin of the new intensities to long-range magnetic
	order. The observation of such a large number of magnetic diffraction peaks
	allowed us to unambiguously determine the magnetic propagation vector, which was
	found to be $\mathbf{k}_1 = (\frac{1}{2}, \frac{1}{2}, 0)$, or $\mathbf{k}_2 =
	(-\frac{1}{2}, \frac{1}{2}, 0)$, or both. Note that these two vectors are
	distinct, i.e. they are not related by an allowed reciprocal lattice vector of
	the C-centred parent structure, yet they cannot be differentiated by powder
	diffraction.
	
	Symmetry analysis was performed using the ISOTROPY Suite [4,5], taking
	the $Cmcm$ crystal structure of CoTi$_2$O$_5$
	\cite{Muller-Buschbaum1983} as the parent. Four irreducible
	representations enter into the decomposition of the magnetic reducible
	representation of $\mathbf{k}_1$ and $\mathbf{k}_2$ for the relevant
	Co Wyckoff positions. Through systematic tests it was found that the
	magnetic structures of just one irreducible representation, mS$_2^{-}$,
	reproduced the relative intensities of the magnetic diffraction
	peaks. In this discussion, we make reference to four
	symmetry-equivalent crystallographic sites, defined with respect to
	the $Cmcm$ unit cell, which comprise the full cobalt sublattice: Co1:
	$[0,y,\frac{1}{4}]$; Co2: $[\frac{1}{2},\frac{1}{2} -y,\frac{3}{4}]$; Co3:
	$[\frac{1}{2},\frac{1}{2} +y,\frac{1}{4}]$; Co4: $[0,1-y,\frac{3}{4}]$, where $y
	= 0.1911(6)$ at 100\,K.
	
	Matrices of the two dimensional irreducible representation mS$_2^{-}$,
	for selected symmetry generators of the parent space group $Cmcm$, are
	given in the top row of Table~S.II \cite{Note2}. The magnetic order parameter
	can take one
	of three distinct directions in the space spanned by the irreducible
	representation: $(\eta, \eta)$, $(\eta, 0)$, or $(\eta, \epsilon)$. In
	all cases, the respective magnetic structures involve moments oriented
	strictly parallel to the orthorhombic $c$ axis. As all of the Co ions
	in the lattice are structurally equivalent, and therefore have the
	same chemical environment, all of the moments on these ions are
	constrained to be equal in magnitude. The only order parameter
	direction consistent with this constraint is
	$(\eta, 0)$.
	
\begin{figure}
		\includegraphics[width=\columnwidth, clip, trim= 0.0mm 0.0mm 0.0mm
		0.0mm]{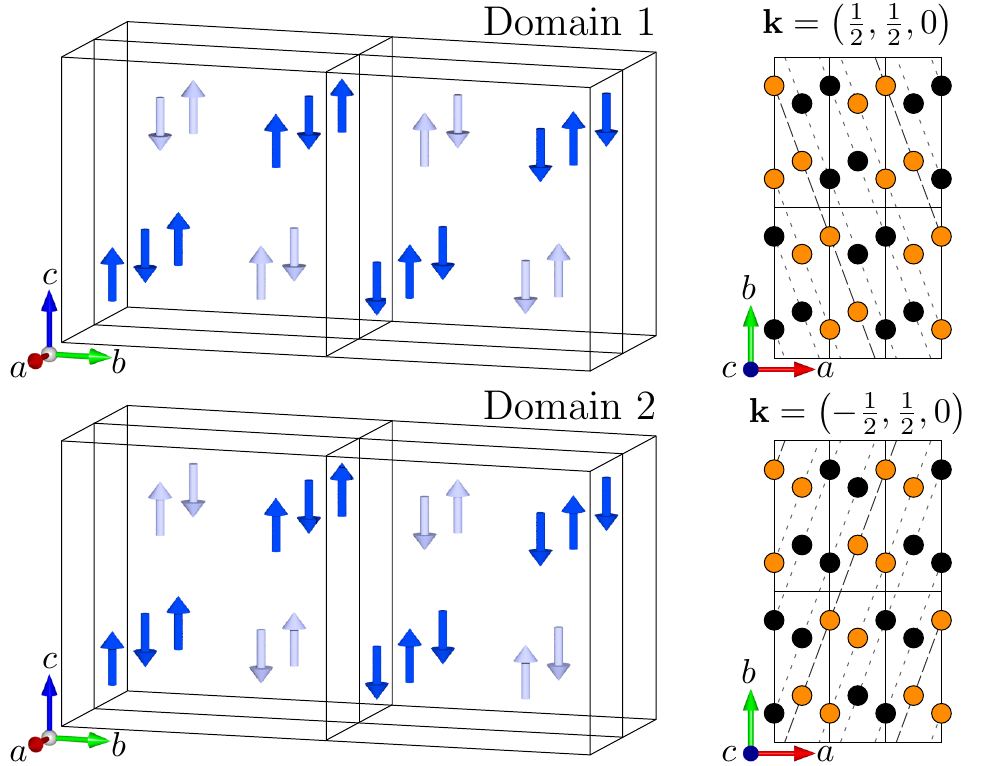}
		\caption{ \label{struct_fig}
			The magnetic structure of CoTi$_2$O$_5$. In the left hand panes four $Cmcm$
			unit cells are drawn, which represent a full repeating unit of the magnetic
			structure. Co1 and Co4 moments, drawn as dark blue arrows, are related to the
			Co2 and Co3 moments, drawn as light blue arrows, by C-centring. Note that the
			two domains related by the $\{m_x|0,0,0\}$ symmetry operator of the parent
			structure can be obtained by reversing the sign of the Co2 and Co3 moments. The
			diagrams on the right hand side illustrate the two propagation vectors
			associated with the two domains, where antiparallel moments are drawn as orange
			and black spheres. Planes of parallel moments are denoted by faint dotted grey
			lines, and the periodicity of the magnetic structure is highlighted by the bold
			dashed grey lines. 
		}
	\end{figure}

	$(\eta, 0)$ corresponds to a magnetic structure that lowers the
	symmetry of the system to monoclinic (magnetic space group $P_a2_1/m$
	\footnote{The $P_a2_1/m$ magnetic
		unit cell has a $\{[-2,0,0],[0,0,1],[\frac{1}{2},\frac{1}{2},0]\}$ change of
		basis
		with respect to the $Cmcm$ parent structure, plus an origin shift of
		$[-\frac{1}{4},\frac{1}{4},0]$. N.B. The orthorhombic $Cmcm$ $c$-axis is
		parallel to the $P_a2_1/m$ $b$-axis in the standard setting.}). Magnetic moments
	on
	the Co1, Co2 and Co4 sites are parallel, but with the moment on the
	Co3 sites aligned antiparallel. A second domain exists with order
	parameter $(0, \eta)$, in which Co1, Co3, and Co4 sites are aligned
	parallel to each other, with Co2 antiparallel. Inspection of the
	mS$_2^{-}$ matrices given in Table~S.II \cite{Note2} shows that the
	$(\eta, 0)$ and $(0, \eta)$ magnetic domains are interchanged by the
	symmetry operator $\{m_x|0,0,0\}$, which is indeed broken below the
	magnetic phase transition. Furthermore, the $(\eta, 0)$ and $(0,
	\eta)$ domains are described by single propagation vectors,
	$\mathbf{k}_1$ and $\mathbf{k}_2$, respectively, which are also
	related by $m_x$. The two domains are shown in
	Fig.~\ref{struct_fig}, with the schematic in the right hand panes
	illustrating the two propagation vector directions. We note that the
	two domains are indistinguishable in our powder diffraction data. The
	magnitude of the Co moment was refined against the diffraction data and found to
	be $2.72(1)\,\mu_{\rm B}$ at 1.5\,K (see Fig.~\ref{npd_fig}(b)).
	
	The only nearest neighbour, super-exchange interactions between cobalt
	atoms (Co--O--Co) connect magnetic moments along the $a$-axis in
	Co$i$--Co$i$ chains. All other nearest-neighbour interactions are
	mediated by super-super-exchange (Co--O--O--Co). One can assume that
	the super-exchange interactions are dominant and, by the
	experimentally determined magnetic structure, are antiferromagnetic.
	All exchange interactions
	between the (Co1,Co4) and (Co2,Co3) sites, coloured dark and light
	blue in Fig.~\ref{struct_fig} respectively, are exactly frustrated by
	the $m_x$ symmetry element. This frustration will likely
	lead to one dimensional ordering of the $a$-axis chains above $T_{\rm
		N}$, but below the mean field energy of the dominant super-exchange
	interaction, consistent with the missing entropy evidenced in the heat capacity.
	For long range order to develop in CoTi$_2$O$_5$, the $m_x$
	mirror symmetry must be broken either at a structural phase transition
	above $T_{\rm N}$, or through the spin Jahn-Teller effect, in which
	the primary magnetic order parameter couples to a secondary, symmetry
	breaking structural order parameter spontaneously at $T_{\rm N}$
	\cite{Yamashita2000, Tchernyshyov2002}. In the absence of any
	experimental evidence for a higher $T$ structural phase
	transition, we discuss possible magneto-structural coupling schemes.
	
	The lowest order, free energy invariant that can couple the magnetic
	order to symmetry breaking crystallographic distortions must be
	quadratic in the magnetic moments (to be time-reversal even), and
	linear in the structural order parameter. On traversing the crystal in
	the direction of the propagation vector, magnetic moments change sign
	from one unit cell to the next. However, in the square of the moments
	each unit cell is the same. Hence, the square of the order parameter
	components, $\eta^2$ and $\epsilon^2$, must couple to a $\mathbf{k}_s
	= (0,0,0)$, $\Gamma$-point structural distortion if the coupling term is to be
	invariant by translation, as required.
	Through exhaustive searches performed using the ISOTROPY suite
	\cite{Campbell2006, Stokes2007}, the only linear-quadratic invariant that can
	couple a non-trivial $\Gamma$-point structural distortion to the magnetic order
	is
	$\delta \left(\eta^2 - \epsilon^2 \right)$,
	where the irreducible representation of the structural order
	parameter, $\delta$, is $\Gamma_{2}^{+}$.
	For completeness, we should also consider the trivial coupling invariant
	$\xi \left(\eta^2 + \epsilon^2 \right)$,
	where the structural order parameter $\xi$ transforms according to
	the totally symmetric $\Gamma_{1}^{+}$ irreducible representation,
	i.e.\ structural distortions that were already allowed within the
	$Cmcm$ parent symmetry can also occur at $T_{\rm N}$.
	The atomic displacements of  $\Gamma_{1}^+$ and 
	$\Gamma_{2}^+$ are tabulated in \cite{Note2}.
	High resolution laboratory based x-ray powder diffraction experiments yielded no
	evidence of these distortions below $T_{\rm N}$. We therefore assume that any
	structural distortion in CoTi$_2$O$_5$ will be small, and the following
	calculations utilize the undistorted unit cell.
	
	\begin{figure}
		\includegraphics[width=\columnwidth, clip, trim= 0.0mm 0.0mm 0.0mm
		0.0mm]{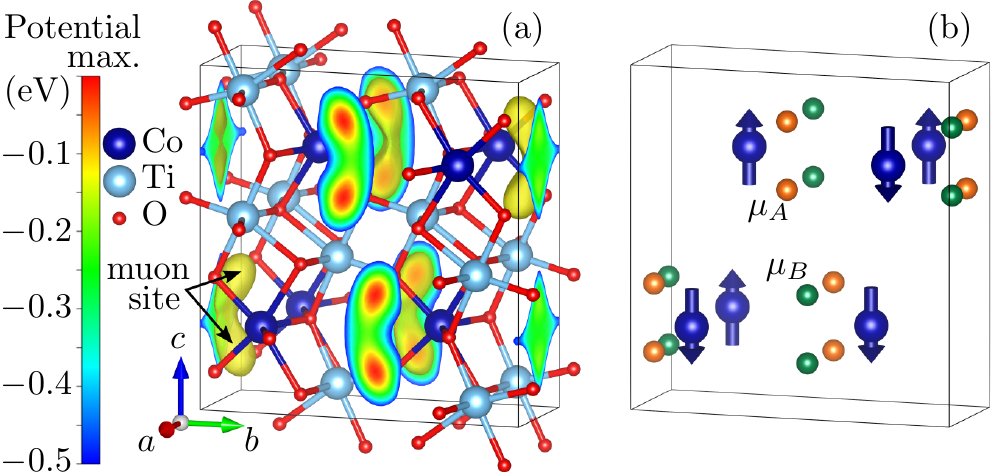}
		\caption{ \label{DFT_fig}
			(a) Electrostatic Coulomb potential of CoTi$_2$O$_5$ computed
			with DFT. The potential is shown on the surface of the unit cell up to $0.5$\,eV
			below its maximum value, and a yellow isosurface is plotted within the unit cell
			at $0.25$~eV below the maximum. (b) Muon positions inside CoTi$_2$O$_5$, with
			the two symmetry inequivalent groups $\mu_A$ and $\mu_B$ marked in orange and
			green respectively. The spin structure of domain 1 is shown on the Co ions. 
		}
	\end{figure}

	In order to establish the potential muon stopping sites in CoTi$_2$O$_5$, we
	employed Density Functional Theory (DFT) calculations to map out the
	electrostatic Coulomb potential of CoTi$_2$O$_5$ throughout its unit cell
	\cite{Note2}, plotted using the VESTA software \cite{Momma2008} in
	Fig~\ref{DFT_fig}(a). The maxima of such a potential map are a reliable
	approximation to the muon sites as they correspond to low energies needed to add
	a positive charge, such as the muon \cite{Moller2013_DFT,Foronda2015}. We also
	carried out relaxation calculations \footnote{See Supplemental Material at [URL
		will be inserted by publisher] for further data fitting and measured parameters,
		details of the DFT calculations, and a discussion of structural order parameters
		and lattice distortions.}, which allow for local distortions of the lattice
	caused by the muon's presence. These gave a single symmetry-inequivalent muon
	stopping site at the general position $[0.322, 0.03, 0.151]$ with a 1.0\,\AA\
	O--H-like bond with the nearest oxygen. This is in line with the approximate
	position we identified from the electrostatic potential.
	
	There are 16 symmetry equivalent muon sites in the $Cmcm$
	parent structure, which are split into two groups of eight in the
	magnetic unit cell, related by the broken $m_x$ symmetry,
	and are denoted by
	$\mu_A$ and $\mu_B$ in Fig.~\ref{DFT_fig}(b). The muon stopping probability
	is dependent upon the electrostatic potential local to the
	stopping sites, and under a small structural distortion induced at the
	phase transition, $\mu_A$ and $\mu_B$ become structurally
	inequivalent, and are therefore associated with different muon
	stopping probabilities. Changing from one magnetoelastic domain to
	another swaps the stopping probabilities of the two subgroups \cite{Note2}. The
	symmetry
	of the magnetic structure also dictates that $\mu_A$ and $\mu_B$ will
	have different local magnetic fields: we calculate these to be
	335(1)\,mT and 277(1)\,mT, in excellent agreement with our
	experimental observations at low $T$. As the area under the
	higher-field peak in Fig.~\ref{ZF_spec}(d) is larger than that of the
	lower-field peak, this suggests that the muon site experiencing this
	field is preferentially occupied. By comparing the energies at the
	muon sites under small distortions, we present a possible coupling
	between a shear distortion and the magnetic domains in \cite{Note2} that could
	explain this.
	
	Finally we consider the additional feature marked by an asterisk in
	Fig.~\ref{ZF_spec}(d) at $\approx 400$\,mT.  This feature likely
	arises due to a Co ion occupying the nearest Ti site so that a small
	fraction of muons stopping close to this defect experience a slightly
	larger field.  Indeed, modelling
	this disorder gives a field at the muon site of $\approx 410$\,mT,
	consistent with the experimental value.
	
	To conclude, we have identified long range magnetic order in
	CoTi$_2$O$_5$, which is antiferromagnetic with $\mathbf{k} = (\pm \frac{1}{2},
	\frac{1}{2}, 0)$. Frustration in the super-super-exchange interactions, along
	with the absence of a structural distortion above $T_{\rm N} \approx
	26$\,K, indicate that the magnetic transition must be coupled to a
	structural transition at $T_{\rm N}$ in order to relieve the
	frustration. This coupling occurs due to the spin Jahn-Teller effect,
	which has so far only been identified in higher-symmetry crystal
	structures \cite{Yamashita2000, Tchernyshyov2002, Onoda2003,
		Watanabe2012}. Our results show that magnetic order driven by
	spin-phonon coupling can be extended to lower-symmetry systems. While
	the predicted distortion in CoTi$_2$O$_5$ was not resolvable in high
	resolution laboratory based x-ray powder diffraction experiments, it
	may be possible to resolve using higher-resolution synchrotron x-ray
	powder diffraction experiments. The study of compounds structurally
	related to CoTi$_2$O$_5$ may provide further insight into the
	conditions required for the spin Jahn-Teller effect to, or not to,
	occur.

	\begin{acknowledgements}
		\emph{Acknowledgements.} F.K.K.K. thanks Lincoln College, Oxford, for a doctoral
		studentship. R.D.J. acknowledges support from a Royal Society University
		Research Fellowship. This work is supported by EPSRC (UK) grant EP/N023803/1.
		Part of this work was performed at the Science and Technology Facilities Council
		(STFC) ISIS Facility, Rutherford Appleton Laboratory, and part at S$\mu$S, the
		Swiss Muon Source (PSI, Switzerland). The authors would like to acknowledge the
		use of the University of Oxford Advanced Research Computing (ARC) facility in
		carrying out this work \cite{Richards2015}.
	\end{acknowledgements}
	
	\bibliographystyle{apsrev4-1}
	\bibliography{bib1}
	
\end{document}